 \input harvmac 
%
\noblackbox 
\def\npb#1#2#3{{\it Nucl.\ Phys.} {\bf B#1} (19#2) #3} 
\def\plb#1#2#3{{\it Phys.\ Lett.} {\bf B#1} (19#2) #3} 
\def\prl#1#2#3{{\it Phys.\ Rev.\ Lett.} {\bf #1} (19#2) #3}

\def\atmp#1#2#3{{\it Adv.\ Theor.\ Math.\ Phys.} {\bf #1} (19#2) #3} 
\def\jhep#1#2#3{{\it JHEP\/} {\bf #1} (19#2) #3} 
\newcount\figno 
\figno=0 
\def\fig#1#2#3{ 
\par\begingroup\parindent=0pt\leftskip=1cm\rightskip=1cm\parindent=0pt 
\baselineskip=11pt 
\global\advance\figno by 1 
\midinsert 
\epsfxsize=#3 
\centerline{\epsfbox{#2}} 
\vskip 12pt 
{\bf Fig.\ \the\figno: } #1\par 
\endinsert\endgroup\par 
} 
\def\figlabel#1{\xdef#1{\the\figno}} 
\def\encadremath#1{\vbox{\hrule\hbox{\vrule\kern8pt\vbox{\kern8pt 
\hbox{$\displaystyle #1$}\kern8pt} 
\kern8pt\vrule}\hrule}} 
 
\def\frac#1#2{{#1 \over #2}}

\def\semi{\subset\kern-1em\times\;} 
\def\bar#1{\overline{#1}} 
%

%

%

\Title{\vbox{\baselineskip12pt 
\hbox{hep-th/0008064} 
\hbox{EFI-2000-26} 
\vskip-.5in}} 
{\vbox{\centerline{Tensionless Branes and Discrete Gauge Symmetry}  
\bigskip}} 
\medskip\bigskip 
\centerline{Jeffrey A. Harvey, Per Kraus, and  Finn Larsen} 
\bigskip\medskip 
\centerline{\it Enrico Fermi Institute and Department of Physics} 
\centerline{\it  University of Chicago,  
Chicago, IL 60637, USA} 
\baselineskip18pt 
\medskip\bigskip\medskip\bigskip\medskip 
\baselineskip16pt 
We argue that the tensionless branes found recently on non-BPS $D$-branes 
using non-commutative field theory are in fact gauge equivalent to the
vacuum under a discrete  gauge symmetry. We also give a simple construction
of the $D(2p)$-branes in IIA theory starting from a single non-BPS $D9$-brane.

\Date{August, 2000} 
\lref\gmsII{
R.~Gopakumar, S.~Minwalla and A.~Strominger,
``Symmetry Restoration and Tachyon Condensation in Open String Theory,''
hep-th/0007226.}
\lref\nati{N.~Seiberg, ``A Note on Background Independence in Noncommutative
Gauge Theories, matrix Model, and Tachyon Condensation,'' hep-th/0008013.}
\lref\wittenflux{E.~Witten,
``Theta Vacua In Two-Dimensional Quantum Chromodynamics,''
Nuovo Cim.\  {\bf A51}, 325 (1979).}
\lref\bsz{N.~Berkovits, A.~Sen and B.~Zwiebach, 
``Tachyon condensation in superstring field theory,'', 
JHEP {\bf 0003}:002, 2000;
hep-th/0002211.} 
\lref\sena{A.~Sen, ``Stable non-BPS bound states of BPS D-branes,'' 
\jhep{9808}{98}{010}, hep-th/9805019;  
``SO(32) spinors of type I and other solitons on brane-antibrane pair,'' 
\jhep{9809}{98}{023}, hep-th/9808141; 
``Type I D-particle and its interactions,'' 
\jhep{9810}{98}{021}, hep-th/9809111;  
``Non-BPS states and branes in string theory,''  
hep-th/9904207, and references therein.} 
\lref\sennon{A. Sen, ``BPS D-branes on non-supersymmetric cycles,''  
\jhep{9812}{98}{021}, hep-th/9812031.} 
\lref\bergman{O.~Bergman and M.~R.~Gaberdiel, 
``Stable non-BPS D-particles,'' \plb{441}{98}{133}, hep-th/9806155.} 
\lref\senuniv{A.~Sen, 
``Universality of the tachyon potential,'' 
JHEP {\bf 9912} (1999) 027, hep-th/9911116.} 
\lref\polchinski{J.~Polchinski, ``Dirichlet-Branes and Ramond-Ramond  
Charges,'' \prl{75}{95}{4724}, hep-th/9510017.} 
\lref\senspinors{A.~Sen, ``$SO(32)$ Spinors of Type I and Other Solitons on  
Brane-Antibrane Pair,'' \jhep{9809}{98}{023}, hep-th/9808141.} 
\lref\ewk{E. Witten, ``D-Branes and K-Theory,'' \jhep{9812}{98}{019};  
hep-th/9810188.} 
\lref\phk{P. Ho\v rava, ``Type IIA D-Branes, K-Theory, and Matrix Theory,''  
\atmp{2}{99}{1373}, hep-th/9812135.} 
\lref\mm{R. Minasian and G. Moore, ``K-Theory and Ramond-Ramond Charge,''  
\jhep{9711}{97}{002}, hep-th/9710230.} 
\lref\yi{P. Yi, ``Membranes from Five-Branes and Fundamental Strings from 
D$p$-Branes,'' \npb{550}{99}{214}; hep-th/9901159.} 
\lref\senpuz{A. Sen, ``Supersymmetric World-volume Action for Non-BPS 
D-branes,'' \hfill\break 
JHEP {\bf 9910}:008, 1999; hep-th/9909062.} 
\lref\senzw{A. Sen and B. Zwiebach, ``Tachyon Condensation in String Field  
Theory,'' hep-th/9912249.} 
\lref\bcr{M. Bill\'o, B. Craps and F.Roose, ``Ramond-Ramond couplings of 
non-BPS D-branes,'' \jhep{9906}{99}{033}; hep-th/9905157.} 
\lref\gms{R. Gopakumar, S.Minwalla and A.Strominger, ``Noncommutative  
Solitons,'' hep-th/0003160.} 
\lref\sw{N. Seiberg and E. Witten, ``String Theory and Noncommutative  
Geometry,'' JHEP {\bf 9909}:032, 1999; hep-th/9908142.} 
\lref\hk{J. A. Harvey and P. Kraus, ``D-Branes as Lumps in Bosonic 
Open String Field Theory,'' JHEP {\bf 0004}:012,2000, hep-th/0002117.} 
\lref\hkm{J. A. Harvey, D. Kutasov and E. Martinec, ``On the Relevance of 
Tachyons,'' hep-th/0003101.} 
\lref\ks{J. Kogut and L. Susskind ``Vacuum Polarization and the  
Absence of Free Quarks in Four-Dimensions,'' Phys. Rev. {\bf D9} 
(1974) 3501.}  
\lref\cds{A.Connes, M. R. Douglas and A.Schwarz, ``Noncommutative Geometry 
and Matrix Theory:Compactification on Tori,'' JHEP {\bf 9802}, 003 (1998), 
hep-th/9711162.} 
\lref\sendes{A. Sen, ``Descent Relations Among Bosonic D-branes,'' 
Int.J. Mod. Phys. {\bf A14} (1999) 4061, hep-th/9902105.} 
\lref\kjmt{R. de Mello Koch, A. Jevicki, M. Mihailescu and R.Tatar, 
``Lumps and P-Branes in Open String Field Theory,'' hep-th/0003031.} 
\lref\bhy{O.Bergman, K.Hori and P.Yi, ``Confinement on the Brane,'' 
hep-th/0002223.} 
\lref\kosts{V. A. Kostelecky and S.Samuel, ``On a Nonperturbative Vacuum for  
the Open Bosonic String,'' Nucl.Phys. {\bf B336} (1990) 263.} 
\lref\mtaylor{N. Moeller and W. Taylor, ``Level Truncation and 
the Tachyon in Open Bosonic String Field Theory,'' hep-th/0002237.} 
\lref\wtaylor{W. Taylor, ``D-brane Effective Field Theory From String 
Field Theory,'' hep-th/0001201.} 
\lref\hkm{J. A. Harvey, D. Kutasov and E. J. Martinec, 
``On the relevance of tachyons'', hep-th/0003101.} 
\lref\bfss{T. Banks, W. Fischler, S.H. Shenker and L. Susskind, 
``M Theory As A Matrix Model: A Conjecture'',  
Phys.Rev.{\bf D55} (1997) 5112, hep-th/9610043.} 
\lref\dwhn{B.~de Wit, J.~Hoppe and H.~Nicolai, 
``On the quantum mechanics of supermembranes'', 
Nucl.\ Phys.\  {\bf B305} (1988) 545.} 
\lref\garousi{M.~R.~Garousi, 
``Tachyon couplings on non-BPS D-branes and Dirac-Born-Infeld action,'' 
hep-th/0003122.} 
\lref\senact{ 
A.~Sen, 
``Supersymmetric world-volume action for non-BPS D-branes,'' 
JHEP {\bf 9910} (1999) 008, 
hep-th/9909062.} 
\lref\bergs{ 
E.~A.~Bergshoeff, M.~de Roo, T.~C.~de Wit, E.~Eyras and S.~Panda, 
``T-duality and actions for non-BPS D-branes,'' 
hep-th/0003221.} 
\lref\kluson{ 
J.~Kluson, 
``Proposal for non-BPS D-brane action,'' 
hep-th/0004106.} 
\lref\dvv{
L.~Motl,
``Proposals on nonperturbative superstring interactions''
hep-th/9701025;
T.~Banks and N.~Seiberg,
``Strings from matrices,''
Nucl.\ Phys.\  {\bf B497} (1997) 41,
hep-th/9702187;
R. Dijkgraaf, E. Verlinde and H. Verlinde, 
``Matrix String Theory'', 
Nucl.Phys. {\bf B500} (1997) 43-61, 
hep-th/9703030.} 
\lref\callm{ 
C.~G.~Callan and J.~Maldacena,  
``Brane Death and Dynamics from the Born-Infeld Action,'' 
Nucl.Phys. {\bf B513} (1998) 198, 
hep-th/9708147.} 
\lref\gibb{G.~W.~Gibbons, 
``Born-Infeld Particles and Dirichlet P-Branes'' 
Nucl.Phys. {\bf B514} (1998) 603, 
hep-th/9709027.} 
\lref\hhk{J.~A.~Harvey, P. Kraus and P. Ho\v rava, 
``D Sphalerons and the Topology of String Configuration Space,'' 
JHEP {\bf 0003}:021,2000, 
hep-th/0001143.}  
\lref\corns{L.Cornalba and R.Schiappa, hep-th/9907211.} 
\lref\ishi{N. Ishibashi, ``A Relation Between Commutative and 
Noncommutative Descriptions of D-Branes,'' hep-th/9909176.} 
\lref\lwone{ 
F.~Larsen and F.~Wilczek, 
``Classical Hair in String Theory I: General Formulation,'' 
Nucl.\ Phys.\  {\bf B475} (1996) 627. 
{hep-th/9604134}.} 
\lref\hlw{P. S.Howe, N.D.Lambert and P.C. West, ``The Selfdual String
Soliton,'' Nucl.Phys. {\bf B515} (1998) 203, hep-th/9709014.}
\lref\dmr{K. Dasgupta, S. Mukhi, and G. Rajesh,
``Noncommutative Tachyons'', hep-th/0005006.}
\lref\malda{J.~Maldacena,
``The large N limit of superconformal field theories and supergravity,''
Adv.\ Theor.\ Math.\ Phys.\  {\bf 2} (1998) 231, hep-th/9711200.}
\lref\luroy{J.X. Lu and S. Roy,
``(p + 1)-Dimensional Noncommutative Yang-Mills and D($p - 2$) Branes'',
hep-th/9912165.}
\lref\sbz{Nathan Berkovits, Ashoke Sen and Barton Zwiebach,
``Tachyon Condensation in Superstring Field Theory'',
hep-th/0002211.}
\lref\wadia{A.~Dhar, G.~Mandal and S.~R.~Wadia,
``String field theory of two-dimensional QCD: 
A Realization of $W_\infty$ algebra,''
Phys.\ Lett.\  {\bf B329} (1994) 15, hep-th/9403050;
``Nonrelativistic fermions, coadjoint orbits of $W_\infty$
and string field theory at $c=1$,''
Mod.\ Phys.\ Lett.\  {\bf A7} (1992) 3129,
hep-th/9207011.}
\lref\hklm{J.~A.~Harvey, P.~Kraus, F.~Larsen and E.~Martinec,
`` D-branes and Strings as Non-commutative Solitons,''
hep-th/0005031.}
\lref\somesenrefb{A. Sen, private communication.}
\lref\somesenrefc{A.~Sen,
``Tachyon condensation on the brane antibrane system,''
JHEP {\bf 9808}, 012 (1998),
hep-th/9805170.}
\lref\somesenrefd{Sen}
\lref\senpriv{Similar constructions were mentioned at A. Sen's talk
at Lennyfest 2000.}
\lref\witten{E. Witten, ``Noncommutative Tachyons and String Field Theory'',
hep-th/0006071.}

It has been shown recently that $D$-branes can be constructed as
exact solitons in open string field theory using techniques
of non-commutative field theory \refs{\gms,\dmr,\hklm,\witten}. 
The resulting solutions have the right tension and spectrum to be
identified with D-branes \hklm . For superstrings one can construct 
BPS Dp-branes as well as non-BPS D-branes as non-commutative 
solitons. All these solitons can be analyzed directly in open string 
field theory \witten. 

These results agree precisely with expectations, but there is an 
additional surprise: 
in type II string theory there are also {\it tensionless} $p$-branes 
\refs{\dmr,\hklm}. If these were genuine light states in type II string 
theory they should have been known already from other studies.
In this note we argue that the tensionless 
solutions are actually gauge transformations of the vacuum and so do 
not appear as new states in the physical spectrum. We use this 
observation to give a new construction of $D(2p)$-branes starting from 
an unstable non-BPS $D9$-brane of type IIA theory.

\newsec{The Brane-anti-Brane System}
Although the emphasis in \refs{\dmr,\hklm} was on solitons on
unstable $Dp$-branes, it is useful to first consider the
$Dp- \overline{Dp}$ system. We start by reviewing this system
in the absence of a background $B$ field and for concreteness start
in IIB theory.  This system has
two gauge fields, $A_+$, $A_-$ and a complex tachyon field $T$ transforming
with charge $(1,-1)$ under $U(1)_+ \otimes U(1)_-$. The gauge transformation
laws are 
\eqn\one{\eqalign{T & \rightarrow U T W^\dagger \cr
                  D_+ & \rightarrow U D_+ U^\dagger \cr
                  D_- & \rightarrow W D_- W^\dagger \cr}}
with $U \in U(1)_+$, $W \in U(1)_-$ and $D_\pm = d + A_{\pm}$.  

We now consider as in \refs{\dmr,\hklm} turning on a background
$B$-field and taking the limit of large non-commutativity. For 
concreteness we consider a B-field in two directions, $B_{89}=B$, 
and take $p=9$; the constructions below are easily generalized to 
more general $B$-fields. In this limit  we can drop ordinary
derivatives in the non-commutative directions, $T$ becomes an arbitrary 
complex operator on Hilbert space ${\cal H}$, and $U, W$ are independent 
unitary transformations on ${\cal H}$. The equations of motion become 
$V'(T)=0$ where operator multiplication is implied \gms. 

We take the classical  tachyon potential $V$ to have a local maximum at
$T=0$ and a ring of minima at $|T| = t_*$.
According to the conjecture of Sen \somesenrefc, $T=0$ represents the
unstable $D9-\overline{D9}$ system and  $|T|=t_*$ represents the closed
string vacuum with no open string excitations. The solution in
the non-commutative theory  \refs{\dmr,\hklm}
\eqn\twoa{T= - t_*(1-P_{k})}
with $P_{k}$ a rank $k$ projection operator on ${\cal H}$
represents $k$ BPS $D7$-branes while the 
tensionless $7$-brane solutions are given by
\eqn\twob{T = -t_* (1-2P_{k})~.}

Choosing a basis of ${\cal H}$ with $P_{k}$ diagonal we
can write the tensionless brane solution as
$T={\rm diag}(t_*^k, -t_*,-t_*, \cdots )$ where
the power on $t_*$ denotes repeated entries, so there are
$k$ entries $+ t_*$. This is clearly gauge equivalent
to the vacuum configuration $T = -t_* 1 $ using the
gauge transformation $U= {\rm diag}(-1^k,1,1,\cdots)$, $W=1$.

\newsec{The non-BPS brane}
We now turn our attention to the unstable $Dp$-branes.
One can obtain a non-BPS $Dp$-brane from a $Dp-{\bar Dp}$ system
by projecting with respect to $(-1)^{F_L}$ \sennon\ . This 
projection sets $A_+=A_-$ and requires that $T$ be real. This breaks 
the  $U(1)_+ \otimes U(1)_-$
gauge symmetry to the subgroup preserving $A_+=A_-$ and the reality of
$T$. This subgroup is $U(1)_c \times Z_2$  where $U(1)_c$ is 
generated by the sum of the generators of $U(1)_+$ and $U(1)_-$
and the $Z_2$ acts as $T \rightarrow -T$. The $Z_2$ symmetry of
the tachyon potential on a non-BPS D-brane in type II can therefore
be viewed as a consequence of a discrete $Z_2$ gauge symmetry
\somesenrefb.

In the non-commutative case we similarly begin with the $D9 - \overline{D9}$ 
system in IIB and project by $(-1)^{F_L}$ to obtain a non-BPS $D9$-brane
in IIA. In analogy to the commutative case, the projection by $(-1)^{F_L}$ 
requires that $A_+=A_-$ and that $T$ be Hermitian. The classical potential 
for $T$ then has a local maximum at $T=0$ representing the unstable 
$D9$-brane and minima at $T= \pm t_*$ representing the closed string vacuum.

In the non-commutative theory, the action is stationary for
$T$ of the form
\eqn\three{T = 0 P_{0} + t_* P_+ - t_* P_-}
where $P_{0},P_+,P_-$ are orthogonal projection operators \gms.
Equivalently, we can diagonalize $T={\rm diag}(t_1,t_2, \cdots)$
and the potential is stationary if each $t_i$ is $0,\pm t_*$.
The stable vacua are represented by
\eqn\aa{T_{vac} = \pm {\rm diag}(-t_*,-t_*,\cdots).}
The discrete $Z_2$ gauge symmetry discussed above interchanges
these two configurations, leaving a single physical vacuum state.
$k$ non-BPS $D7$-branes are represented by
\eqn\bb{T_{D7}=\pm(0^k,-t_*,-t_*,\cdots)}
with the two solutions again interchanged by $Z_2$. The final 
solutions of interest are the tensionless $7$-branes, given by
\eqn\cc{T_{ten}=\pm(t_*^k,-t_*,-t_*,\cdots).}
We would like to show that $T_{ten}$ is gauge equivalent to 
$T_{vac}$, but note that this does not follow from the above
$Z_2$ gauge symmetry. 

Instead, we require a gauge symmetry allowing us to flip the
eigenvalues of $T$ independently. The need for such a symmetry
can be understood on the following grounds.  Consider a
fluctuation of the tachyon on a single non-BPS $D7$-brane,
\eqn\dd{T_{D7}+\delta T_{D7} = \pm(\delta t,-t_*,-t_*,\ldots).}
Since we can construct the non-BPS $D7$-brane as the $(-)^{F_L}$
projection of the $D7-\bar{D7}$ system, it follows that there must 
be a $Z_2$ gauge symmetry flipping the sign of the $D7$ tachyon,
$\delta t \rightarrow -\delta t$.  Generalizing to arbitrary 
numbers of $D7$-branes, it follows that there must exist a 
gauge symmetry allowing us to flip the sign of any collection 
of eigenvalues, and under this symmetry $T_{ten}$ is gauge 
equivalent to $T_{vac}$.  Therefore the tensionless solutions
found in \refs{\dmr,\hklm} with $T={\rm diag}(t_*^k,-t_*,-t_*,\cdots)$
are gauge equivalent to the vacuum and should not be counted as 
distinct solutions.
Similarly, we can show that a superposition of $k$ non-BPS $D7$-branes
and $k'$ tensionless $7$-branes is gauge equivalent to k non-BPS $D7$-branes.
We can always use the the Weyl group of $U(\infty)$ to put
the tachyon
field \three\ into the form $T={\rm diag} (0^{n_0},-t_*^{n_-},t_*^{n_+})$ 
with $n_0+n_-+n_+$ infinite.  For $n_0$
finite this solution has the same tension and spectrum as 
$n_0$ non-BPS $D7$-branes \hklm.   By viewing this solution as a
tachyon configuration on $n_0 + n_-$ $ D7$-branes we can use the $Z_2$
symmetry of the tachyon on these $D7$-branes to map this solution to
the canonical form $T={\rm diag} (0^{n_0},t_*^{n_-},t_*^{n_+})$. 

Let us now try to identify this $Z_2$ gauge symmetry directly.
Setting $T=T^\dagger$ and $A_+ = A_-$ in the action of the 
non-commutative $D9-\bar{D9}$ system gives
\eqn\ee{S=\int \! d^8x \, {\rm Tr}\left\{ D_\mu T D^\mu T
- {1 \over 4} F_{\mu\nu}F^{\mu\nu} -V(T)\right\}
}%
with $D_\mu T = \partial_\mu T + i[A_\mu,T]$.  Due to the  
$[A_\mu,T]$ terms, it is clear that for generic $A_\mu$ 
flipping the sign of a $T$ eigenvalue is {\it not} a symmetry 
of the action --- trouble comes from the off-diagonal elements
of $A_\mu$. 

As discussed in \hklm, when expanding the action around the
background of $k$ $D7$-branes the gauge bosons of
$U(\infty)/(U(\infty-k) \times U(k))$ appear as unwanted massive
degrees of freedom. Their mass cannot be computed reliably in an
effective field theory approach, and it was argued that
they should be removed by higher derivative terms in the full
string field theory.  Similarly,
here we propose that when expanding around a $7$-brane solution
it is necessary to freeze out these off-diagonal degrees of
freedom by setting them to zero.  Indeed, this is necessary in
order to recover the $T \rightarrow -T$ gauge symmetry on a 
non-BPS $D7$-brane.  Of course, it would be desirable to see
this happening explicitly, but given this physically 
well-motivated assumption we have 
shown that the tensionless solutions are gauge equivalent to 
the vacuum.

We also note that additional brane and vacuum solutions
have appeared in \refs{\gmsII,\nati}.
One would
like to find a formulation in which the $D$-brane solutions and
vacuum are unique and any additional solutions are gauge artifacts,
as we have found here for the tensionless branes in type II theory.

\newsec{A Construction of $D(2p)$-branes}
We can also  construct non-trivial solutions which interpolate
between vacua related by discrete gauge transformations. For example,
a BPS $D8$-brane is represented in the commutative theory 
by a kink which interpolates between
the vacua at $T= \pm t_*$. In the present context we can consider
a tachyon field which also depends on one of the commuting directions,
say $x_7$, in which case a $D8$ (anti-$D8$)-brane would be given by the
configuration
\eqn\four{T = t_* \Phi_{k, (\bar k)}(x_7)}
with $\Phi_{k,(\bar k)}(x_7)$ a 
kink (anti-kink) configuration which interpolates
between $\mp (\pm) 1$ as $x_7$ varies from $- \infty$ to $+ \infty$.
Since this configuration depends on a commuting coordinate, it
is not possible to compute its tension exactly as in \hklm. It
is nonetheless clear that it represents a $D8$-brane
(with a $B$-field in two directions along the brane). 

We can also construct $D6$-branes using these ideas in terms of
solutions that  interpolate
between vacua where a finite number of eigenvalues differ in sign \senpriv.
Take $P_\pm$ to be orthogonal projection operators of rank $n_+,n_-$.
Then the solution
\eqn\five{T = t_*\Phi_k(x_7)P_+ + 
t_*\Phi_{\bar k}(x_7)P_- + t_*(1-P_+ - P_-)}
represents a superposition of $n_+$ $D6$-branes and $n_-$ anti
$D6$-branes. To see this, note that the solution with $n_++n_-$
zeroes on the diagonal and the remaining diagonal entries equal to
$t_*$ represents $n_+ + n_-$ non-BPS
$D7$-branes. $D6$-branes are represented by kinks on a non-BPS
$D7$-brane, and the above construction has $n_+$ kinks and $n_-$
anti-kinks in commuting subspaces. This construction should be 
contrasted with the construction of $D6$-branes as 't 
Hooft-Polyakov monopoles on several $D9$-branes in the commutative 
framework \phk. In \five\ only a single $D9$-brane is required
and only the tachyon
is excited, whereas in \phk\ the gauge field is essential. 

Similarly, by turning on a B-field in more 
directions, one can construct all the BPS D(2p)-branes of IIA
as generalized kinks on a single non-BPS D9-brane.
 
\bigskip\medskip\noindent 
{\bf Acknowledgements:} 
This work was supported in part by NSF grant PHY-9901194 and by DOE grant
DE-FG0290ER-40560. J.H. and P.K thank the Aspen Center for Physics for 
hospitality during the completion of this work. FL was supported
in part by a Robert R. McCormick fellowship. 
 
\listrefs 
\end